\documentclass[10pt, final, journal, letterpaper, oneside, twocolumn]{IEEEtran}
\usepackage{graphicx}
\usepackage{amsmath}
\usepackage{amssymb}
\usepackage{mathrsfs}
\usepackage{stfloats}
\usepackage{dsfont}
\usepackage{setspace}
\usepackage{array}
\usepackage{bbm}
\usepackage{caption}
\usepackage{subcaption}
\usepackage{algorithmic}
\usepackage{algorithm}
\usepackage{epstopdf}
\usepackage{xcolor}
\usepackage{accents}

\newtheorem{theorem}{\bf{Theorem}}

\newtheorem{proposition}{\bf{Proposition}}

\newtheorem{remark}{\bf{Remark}}

\DeclareMathAlphabet\mathbfcal{OMS}{cmsy}{b}{n}
\begin{document}
\title{Mutual Coupling and Unit Cell Aware Optimization for Reconfigurable Intelligent Surfaces}
\author{Xuewen~Qian and Marco~Di~Renzo, \IEEEmembership{Fellow,~IEEE} %\vspace{-0.75cm}
\thanks{Manuscript received Nov. 29, 2020. X. Qian and M. Di Renzo are with CNRS and Paris-Saclay University, France (e-mail: marco.di-renzo@universite-paris-saclay.fr).}
}
%\markboth{Wireless Communications Letters} {X. Qian and M. Di Renzo, Mutual Coupling and Unit Cell Aware Optimization of Reconfigurable Intelligent Surfaces}
%
%
%
%
\maketitle
\begin{abstract}
Reconfigurable intelligent surfaces (RISs) are an emerging technology for enhancing the performance of wireless networks at a low and affordable cost, complexity, and power consumption. We introduce an algorithm for  optimizing a single-input single-output RIS-assisted system in which the RIS is modeled by using an electromagnetic-compliant framework based on mutual impedances. More precisely, we provide the following new contributions: (i) in the absence of mutual coupling among the scattering elements of the RIS, we derive a closed-form expression for the optimal tunable impedances, which inherently accounts for the interplay between the amplitude and phase of the lumped loads of the RIS; and (ii) in the presence of mutual coupling, we introduce an iterative algorithm for optimizing the tunable impedances of the RIS. The algorithm is proved to be convergent by showing that the objective function is non-decreasing and upper bounded. Numerical results reveal that the mutual coupling among the scattering elements of the RIS significantly affects the end-to-end signal-to-noise ratio (SNR) if the inter-distance is less than half of the wavelength. If the RIS is optimized by explicitly taking into account the impact of mutual coupling, a better end-to-end SNR is obtained.
\end{abstract}
%
%f
%
%
%

\section{Introduction} \label{Introduction} 
A reconfigurable intelligent surface (RIS) is an emerging technology that enables the control of the electromagnetic waves at a reduced cost, power consumption, and hardware complexity \cite{MDR_JSAC}. In general terms, an RIS can be viewed as a thin sheet of electromagnetic material, which is made of a large number of nearly-passive scattering elements that are controlled via low cost and low power electronic circuits. By appropriately configuring the electronic circuits, different wave transformations can be applied. Recent research works have shown that RISs whose geometric size is sufficiently large are capable of outperforming other technologies, e.g., relays, at a reduced hardware and signal processing complexity \cite{MDR_Relays}, and are capable of enhancing the reliability of wireless links by reducing the fading severity \cite{Xuewen_scalingLaw}. In addition, the achievable performance of RIS-assisted systems has been proved to be robust to various hardware impairments, e.g., the phase noise, which may further reduce the implementation cost \cite{MDR_Diversity}.

In order to quantify the performance gains offered by RISs in wireless networks, realistic communication models need to be employed, see, e.g., \cite{RuiZhang} and \cite{Gabriele}. The authors of \cite{Gabriele}, in particular, have recently introduced an end-to-end communication model for RIS-assisted systems that accounts for the mutual coupling among the closely spaced scattering elements of the RIS, and for the circuits of the electronic components that are used for making the RIS reconfigurable. The communication model proposed in \cite{Gabriele} is derived by departing from Maxwell's equations and leverages a general framework based on self and mutual impedances. More precisely, an RIS is modeled as an array of passive scattering elements driven by low power tunable lumped impedances, which can be appropriately optimized to control the scattered electromagnetic waves.

By capitalizing on the impedance-based communication model in \cite{Gabriele}, we introduce an analytical and numerical framework for optimizing the tunable lumped impedances of an RIS, so as to maximize the end-to-end signal-to-noise ratio (SNR). In particular, we focus our attention on a single-input single-output RIS-assisted system and provide the following contributions: (i) in the absence of mutual coupling among the scattering elements of the RIS, we derive a closed-form expression for the optimal tunable lumped impedances, which inherently accounts for the interplay between their amplitude and phase response; and (ii) in the presence of mutual coupling, we introduce an iterative algorithm for optimizing the tunable lumped impedances. The algorithm is shown to be convergent by proving that the objective function is non-decreasing and upper bounded. Numerical results reveal that the mutual coupling among the scattering elements of the RIS significantly affects the end-to-end SNR if they are spaced less than half of the wavelength apart. If the RIS is optimized by explicitly taking into account the mutual coupling, a better end-to-end SNR is attained.

The rest of this paper is organized as follows. In Section II, the system model and the problem statement are introduced. In Sections III and IV, the optimization frameworks in the absence and in the presence of mutual coupling are presented, respectively. In Section V, numerical results are illustrated and discussed. Finally, Section VI concludes this paper.

\textit{Notation}: Vectors and matrices are denoted in bold font; $j=\sqrt{-1}$ denotes the imaginary unit; $ \textrm{Re}(a)$ and $\textrm{Im}(a)$ denote the real and imaginary parts of the complex number $a$, respectively; $| a | $ and $\angle a$ denote the absolute value and the phase of the complex number $a$, respectively; ${{\bf{A}}^H}$ denotes the Hermitian of matrix $\bf{A}$; ${w_\pi }\left( \theta  \right)$ is the function that wraps the angle $\theta$, expressed in radians, to the interval $\left[ { - \pi , + \pi } \right]$; ${{\bf{0}}_{N \times N}}$ denotes an $N \times N$ matrix whose entries are all zero; ${{\bf{I}}_{N \times N}}$ denotes an $N \times N$ identity matrix; $\| \bf{A} \|$ denotes the spectral norm of $\bf{A}$ that is equal to the largest eigenvalue of ${{\bf{A}}^H}{\bf{A}}$.

\section{System Model and Problem Formulation} \label{System Model}

We consider an RIS-assisted wireless system that comprises a single-antenna transmitter, a single-antenna receiver, and an RIS that is made of $N_{\rm{RIS}} = M \times M$ scattering elements arranged on a square array. The locations of the transmitter $t$, the receiver $r$, and the $i$th passive scatterer of the RIS are denoted by ${{\bf{r}}_{\xi}} = {x_{\xi}}{\bf{\hat x}} + {y_{\xi}}{\bf{\hat y}} + {z_{\xi}}{\bf{\hat z}}$ for ${\xi} = \{ t,r,i\}$. According to \cite{Gabriele}, we assume that the transmit antenna, the receive antenna, and the scattering elements of the RIS can be modeled as cylindrical thin wires of perfectly conducting material, whose length is $l$ and whose radius $a \ll l$ is finite but negligible with respect to $l$ (thin wire regime). The inter-distance between adjacent scattering elements of the RIS is denoted by $d$. Since the transmitter, the receiver, and the RIS may have different implementation requirements, we consider distinct pairs $(l_{\xi}, a_{\xi})$. All thin wires are assumed to be parallel to each other and to be co-polarized, as shown in \cite[Fig. 1]{Gabriele}. 

Each scattering element of the RIS is connected to a lumped load impedance, which can be optimized in order to appropriately shape the propagation of the electromagnetic waves. The $N_{\rm{RIS}}$ tunable impedances are collected in the $N_{\rm{RIS}} \times N_{\rm{RIS}}$ diagonal matrix $\mathbf{Z}_{\rm{RIS}}$, whose $i$th diagonal element is denoted by $\mathbf{Z}_{\rm{RIS}}(i,i)$. In particular, ${\mathop{\rm Re}\nolimits} \left( {{{\bf{Z}}_{{\rm{RIS}}}}\left( {i,i} \right)} \right) = {R_0} \ge 0$, for $i=1, 2, \ldots, N_{\rm{RIS}}$, denotes the resistance of each lumped load. The resistance $R_0$ accounts for the internal losses of the tuning circuits and is assumed to be fixed. On the other hand, the reactance ${\mathop{\rm Im}\nolimits} \left( {{{\bf{Z}}_{{\rm{RIS}}}}\left( {i,i} \right)} \right)$ is an arbitrary real number that can be appropriately optimized for system optimization.

By assuming that the transmitter and the receiver are in the far-field of each other and of the RIS, the end-to-end channel of the considered RIS-assisted system can be formulated as:
\begin{eqnarray}
\mathcal{H}_{\rm{E2E}} = \mathcal{Y}_{0} (Z_{\rm{RT}}-\mathbf{z}_{\rm{RS}} (\mathbf{Z}_{\rm{SS}}+\mathbf{Z}_{\rm{RIS}} )^{-1}\mathbf{z}_{\rm{ST}} ) \label{eq:He2e_SISO}
\end{eqnarray}
\noindent where $\mathcal{Y}_{0}$ is a complex constant that accounts for the internal impedance of the voltage generator at the transmitter, the load impedance at the receiver, the self impedances of the transmit and receive antennas; $Z_{\rm{RT}}$ is the mutual impedance between the transmitter and the receiver; $\mathbf{z}_{\rm{ST}}$ is the $N_{\rm{RIS}} \times 1$ vector of mutual impedances between the transmit antenna and the reconfigurable elements of the RIS; $\mathbf{z}_{\rm{ST}}$ is the $1 \times N_{\rm{RIS}}$ vector of mutual impedances between the reconfigurable elements of the RIS and the receive antenna; and  $\mathbf{Z}_{\rm{SS}}$ is the $N_{\rm{RIS}} \times N_{\rm{RIS}}$ matrix of self and mutual impedances between pairs of reconfigurable elements of the RIS. The mutual impedances can be computed by using \cite[Lemma 2]{Gabriele}, and they only depend on the geometry of
the considered thin wire antennas.

In this paper, we are interested in optimizing the diagonal matrix of tunable impedances $\mathbf{Z}_{\rm{RIS}}$ in order to maximize the intensity of the voltage measured at the port of the receiver. Based on the end-to-end channel in \eqref{eq:He2e_SISO}, the corresponding optimization problem can be formulated as follows:
\begin{subequations}
\begin{align}
\mathcal{P}: \quad & \max_{\mathbf{Z}_{\rm{RIS}}} |Z_{\rm{RT}}-\mathbf{z}_{\rm{RS}} (\mathbf{Z}_{\rm{SS}}+\mathbf{Z}_{\rm{RIS}} )^{-1}\mathbf{z}_{\rm{ST}} |\label{eq:optimization_problem_SISO}\\
 \mathrm{subject \; to} & \quad \textrm{Re} (\mathbf{Z}_{\rm{RIS}}(i,i) ) = R_0 \ge 0 \label{eq:optimization_problem_SISO_constraint_1} \\
& \quad \textrm{Im} (\mathbf{Z}_{\rm{RIS}}(i,i) ) \in \mathbb{R}
\label{eq:optimization_problem_SISO_constraint_2}
\end{align}
\end{subequations}

In the following two sections, we solve $\mathcal{P}$ under the assumption that the mutual coupling among the scattering elements of the RIS can be ignored and cannot be ignored, respectively.

\section{Optimization -- No Mutual Coupling}\label{sec:No_Coupling}

In this section, we assume that no mutual coupling among the scattering elements of the RIS exists. Therefore, $\mathbf{Z}_{\rm{SS}}$ is a diagonal matrix and \eqref{eq:He2e_SISO} simplifies to:
\begin{align}
\mathcal{H}_{\rm{E2E}} & =\mathcal{Y}_{0}
\left( \mathbf{\mathrm{Z}}_{\rm{RT}}-\sum_{i=1}^{N_{\rm{\rm{RIS}}}}\frac{\mathbf{z}_{\rm{ST}}(i)\mathbf{z}_{\rm{RS}}(i)}{\mathbf{Z}_{\rm{SS}}(i,i)+\mathbf{Z}_{\rm{RIS}}(i,i)} \right)\label{eq:He2e_SISO_no_coupling}
\end{align}

In the considered system model, the $N_{\rm{RIS}}$ scattering elements of the RIS are identical. Therefore, ${{\bf{Z}}_{{\rm{SS}}}}\left( {i,i} \right) = {X_{{\rm{SS}}}} + j{Y_{{\rm{SS}}}}$ for $i=1,2,\ldots,N_{\rm{RIS}}$. For ease of writing, we introduce the notation $\mathbf{Z}_{\rm{RIS}}(i,i) + \mathbf{Z}_{\rm{SS}}(i,i) = x_{i}+jy_{i}$, where:
\begin{align} \label{Eq_Phase_1}
& {x_i} = {\mathop{\rm Re}\nolimits} \left( {{{\bf{Z}}_{{\rm{RIS}}}}\left( {i,i} \right) + {{\bf{Z}}_{{\rm{SS}}}}\left( {i,i} \right)} \right) = {R_0} + {X_{{\rm{SS}}}}\\
& {y_i} = {\mathop{\rm Im}\nolimits} \left( {{{\bf{Z}}_{{\rm{RIS}}}}\left( {i,i} \right) + {{\bf{Z}}_{{\rm{SS}}}}\left( {i,i} \right)} \right) = {\mathop{\rm Im}\nolimits} \left( {{{\bf{Z}}_{{\rm{RIS}}}}\left( {i,i} \right)} \right) + {Y_{{\rm{SS}}}} \nonumber 
\end{align}

Therefore, we obtain:
\begin{align} \label{Eq_Phase_2}
\frac{1}{{{x_i} + j{y_i}}} = \frac{{{x_i} - j{y_i}}}{{x_i^2 + y_i^2}} = {\rho _i}\exp \left( {j{\theta _i}} \right)
\end{align}
\noindent where:
\begin{align} \label{Eq_Phase_3}
\tan \left( {{\theta _i}} \right) =  - \frac{{{y_i}}}{{{x_i}}}, \quad {\rho _i} = \frac{1}{{\sqrt {x_i^2 + y_i^2} }} = \frac{{\left| {1 + \exp \left( {j2{\theta _i}} \right)} \right|}}{{2\left| {{x_i}} \right|}}
\end{align}

Since the phase of ${1 + \exp \left( {j2{\theta _i}} \right)}$ is equal to $\theta_i$, i.e., $\angle \left( {1 + \exp \left( {j2{\theta _i}} \right)} \right) = {\theta _i}$, we have $1 + \exp \left( {j2{\theta _i}} \right) = \left| {1 + \exp \left( {j2{\theta _i}} \right)} \right|\exp \left( {j{\theta _i}} \right)$. From \eqref{Eq_Phase_1}-\eqref{Eq_Phase_3}, therefore, we obtain:
\begin{align} \label{Eq_Phase_Final}
\frac{1}{{{x_i} + j{y_i}}} = {\rho _i}\exp \left( {j{\theta _i}} \right) = \frac{{1 + \exp \left( {j2{\theta _i}} \right)}}{{2\left| {{x_i}} \right|}}
\end{align}

\begin{remark}
It is worth mentioning that ${\phi _i} = 2{\theta _i} \in \left[ {- \pi,\pi } \right)$ for any values of $X_{\rm{SS}}$, since the reactance $Y_{\rm{SS}}$ can be positive and negative. If $X_{\rm{SS}} \ge 0$, in particular, we have $x_i \ge 0$, which implies ${\theta _i} \in \left[ {-\pi/2,\pi/2 } \right)$ and ${\phi _i} \in \left[ {-\pi,\pi } \right)$. 
\end{remark}

With the aid of \eqref{eq:He2e_SISO_no_coupling} and \eqref{Eq_Phase_Final}, $\mathcal{P}$ can be reformulated as:
\begin{subequations}
\begin{align}
\mathcal{P_{\rm{NC}}}: \quad & \mathop {\max }\limits_{\scriptstyle{\theta _i}\atop
\scriptstyle i = 1,2, \ldots ,{N_{{\rm{RIS}}}}} \left| {b - \sum\limits_{i = 1}^{{N_{{\rm{RIS}}}}} {{a_i}\exp \left( {j2{\theta _i}} \right)} } \right| \label{eq:eq:He2e_SISO_NC_max_H_v}\\
 \mathrm{subject \; to} & \quad 2{\theta _i} \in \left[ {-\pi,\pi } \right), \quad i= 1,2,\ldots, {N_{{\rm{RIS}}}}
\end{align}
\end{subequations}
\noindent where the following definitions hold:
\begin{align} 
{a_i} = \frac{{{{\bf{z}}_{{\rm{ST}}}}\left( i \right){{\bf{z}}_{{\rm{RS}}}}\left( i \right)}}{{2\left| {{R_0} + {X_{{\rm{SS}}}}} \right|}},\quad b = {Z_{{\rm{RT}}}} - \sum\limits_{i = 1}^{{N_{{\rm{RIS}}}}} {{a_i}}
\end{align}

Let $\theta _i^*$ for $i = 1,2, \ldots ,{N_{{\rm{RIS}}}}$ be the solution of $\mathcal{P_{\rm{NC}}}$. The optimal tunable impedance of the $i$th element of the RIS is:
\begin{align} \label{Impedance_SISO}
{\bf{Z}}_{{\rm{RIS}}}^*\left( {i,i} \right) = \frac{{2\left| {{R_0} + {X_{{\rm{SS}}}}} \right|}}{{1 + \exp \left( {j2\theta _i^*} \right)}} - {{\bf{Z}}_{{\rm{SS}}}}\left( {i,i} \right)
\end{align} 

The optimal phase shifts $\theta _i^*$ for $i = 1,2, \ldots ,{N_{{\rm{RIS}}}}$ in \eqref{eq:eq:He2e_SISO_NC_max_H_v} are given in the following theorem.

\begin{theorem} \label{thm:SISO_NC_theorem}
The global maximizer of $\mathcal{P_{\rm{NC}}}$ is:
\begin{equation}
2\theta_i^* =  {w_\pi }(\angle b - \angle  a_i \pm \pi ) \label{eq:v_theory_SISO_NC}
\end{equation}

Also, the corresponding intensity of the end-to-end channel $\mathcal{H}_{\rm{E2E}}$ is the following:
\begin{equation}
\left| {{{\mathcal{H}}_{{\rm{E2E}}}}} \right| = \left| {{{\mathcal{Y}}_0}} \right|\left| {\left|b\right| + \sum\limits_{i = 1}^{{N_{{\rm{RIS}}}}} {\left| {{a_i}} \right|} } \right|
\end{equation}
\end{theorem}
\begin{IEEEproof}
Define the objective function $F\left( {\boldsymbol {\theta }} \right)$, where $\boldsymbol \theta$ is the vector of phase shifts to be optimized:
\begin{equation}
    F\left( {\boldsymbol{\theta }} \right) = \left| {b - \sum\limits_{i = 1}^{{N_{{\rm{RIS}}}}} {{a_i}\exp \left( {j2{\theta _i}} \right)} } \right|
\end{equation}

The stationary points of $F\left( {\boldsymbol{\theta }} \right)$ correspond to the solution of the set of equations ${{\partial F\left( {\boldsymbol{\theta }} \right)} \mathord{\left/
{\vphantom {{\partial F\left( {\boldsymbol{\theta }} \right)} {\partial {\theta _i}}}} \right. \kern-\nulldelimiterspace} {\partial {\theta _i}}} = 0$ for $i = 1,2, \ldots ,{N_{{\rm{RIS}}}}$, which can be formulated as follows:
\begin{align} \label{StationaryPoints}
\left| {b{a_i}} \right|\sin \left( {{\chi _i}} \right) - \sum\limits_{k \ne i = 1}^{{N_{{\rm{RIS}}}}} {\left| {{a_i}{a_k}} \right|\sin \left( {{\chi _i} - {\chi _k}} \right)}  = 0
\end{align}
\noindent where ${\chi _i} = \angle \beta  - \angle {a_i} - 2{\theta _i}$ for $i = 1,2, \ldots ,{N_{{\rm{RIS}}}}$.

The zeros of \eqref{StationaryPoints} are ${\chi _i} = 0$ and ${\chi _i} = \pm \pi$ for $i = 1,2, \ldots ,{N_{{\rm{RIS}}}}$. The proof follows by noting that $F\left( {\boldsymbol {\theta }} \right)$ attains its maximum if ${\chi _i} = \pm \pi$ for $i = 1,2, \ldots ,{N_{{\rm{RIS}}}}$. This can be proved by applying the triangle inequality:
\begin{align} \label{OptimizedSISO}
F\left( {\boldsymbol{\theta }} \right) = \left| {b - \sum\limits_{i = 1}^{{N_{{\rm{RIS}}}}} {{a_i}\exp \left( {j2{\theta _i}} \right)} } \right| \le \left| b \right| + \sum\limits_{i = 1}^{{N_{{\rm{RIS}}}}} {\left| {{a_i}} \right|}
\end{align}
\noindent where the upper bound is attained by setting $\beta  - \angle {a_i} - 2{\theta _i} = \pm \pi$ for $i = 1,2, \ldots ,{N_{{\rm{RIS}}}}$. This completes the proof.
\end{IEEEproof}

\begin{remark}
The optimization of the impedance-based end-to-end channel model for RIS-assisted systems is different from the usual optimization of single-input single-output RIS-assisted systems, e.g., \cite{Xuewen_scalingLaw} and \cite{QingqingWu_scalingLaw}. More precisely, by direct inspection of \eqref{Impedance_SISO}, we observe that: (i) the amplitude and the phase of the tunable load impedances are not independent of each other; and (ii) the internal losses (through $R_0$) of the tunable circuits are explicitly taken into account along with the self impedance of the radiating scattering elements (through ${{{\boldsymbol{Z}}_{{\rm{SS}}}}}$). These two remarks are similar to those made in \cite{RuiZhang} for a different circuit model of the RIS. In contrast to \cite{RuiZhang}, Theorem \ref{thm:SISO_NC_theorem} yields the closed-form and globally optimal expression of the tunable impedances that maximize the end-to-end SNR. This is a major positive aspect of the considered impedance-based model for RIS-assisted communications.
\end{remark}

\begin{remark}
By direct inspection of \eqref{OptimizedSISO}, we evince that, in the far-field regime, the SNR of an optimized RIS scales with the square of the number of reconfigurable scattering elements, i.e., ${\rm{SNR}} \propto {\left| {F\left( {\boldsymbol{\theta }} \right)} \right|^2} \propto N_{{\rm{RIS}}}^2$. This is in agreement with conventional models for RIS-assisted communications, e.g., \cite{Xuewen_scalingLaw} and \cite{QingqingWu_scalingLaw}. In addition, the SNR of the signal scattered by an RIS scales with the reciprocal of the product of the square of the transmission distances between the transmitter and the mid-point of the RIS, and the mid-point of the RIS and the receiver. This is similar to \cite{Wankai} and \cite{MDR_PathLoss}.
\end{remark}

\section{Optimization -- Mutual Coupling}

In this section, we analyze the general setup in which the mutual coupling among the reconfigurable elements of the RIS cannot be ignored. In this case, the off-diagonal elements of $\mathbf{Z}_{\rm{SS}}$ are not negligible, and $\mathbf{Z}_{\rm{SS}}$ can not be approximated to be a diagonal matrix. Compared with the case study in the absence of mutual coupling, solving $\mathcal{P}$ entails the inversion of $\mathbf{Z}_{\rm{SS}} + \mathbf{Z}_{\rm{RIS}}$, i.e., the computation of $\left(\mathbf{Z}_{\rm{SS}} + \mathbf{Z}_{\rm{RIS}}\right)^{-1}$. To circumvent this issue, we propose an iterative, provably convergent, algorithm that capitalizes on the optimal solution obtained in the absence of mutual coupling.

\begin{algorithm}[t!]\label{alg:Overal_algorithm}
\begin{algorithmic}[1]
\STATE \textbf{initialization}: $\mathbf{Z}_{\rm{RIS}}^{(1)}=\mathbf{Z}_{\rm{RIS}}^{\rm{NC}}$, $\mathbf{Z}_{D}^{(1)}={{\bf{0}}_{{N_{{\rm{RIS}}}} \times {N_{{\rm{RIS}}}}}}$
\STATE \textbf{for} $k=1:1:K$
\STATE \quad{}Solve $\mathcal{P}_{D}^{(k)}$
to obtain $\mathbf{Z}_{D}^{(k)*}$
\STATE \quad{}Update $\mathbf{Z}_{\rm{RIS}}^{(k)}$ as in \eqref{eq:updation_Z_ris_k_plus_1}
\STATE \textbf{end}
\STATE \textbf{return}  $\mathbf{Z}_{\rm{RIS}}^{(K)}$ 
\end{algorithmic} 
\caption{\label{alg:Nt1Nr1_coupling_overall_procedures} Proposed optimization algorithm} 
\end{algorithm} 
The proposed algorithm is reported in Algorithm \ref{alg:Nt1Nr1_coupling_overall_procedures}. The following notation is used: (i) $\mathbf{Z}_{\rm{RIS}}^{(k)}$ denotes the value of $\mathbf{Z}_{\rm{RIS}}$ at the $k$th iteration; (ii) $\mathbf{Z}_{\rm{RIS}}^{\rm{NC}}$ is the optimal matrix of tunable impedances of the RIS in the absence of mutual coupling. $\mathbf{Z}_{\rm{RIS}}^{\rm{NC}}$ is obtained from \eqref{Impedance_SISO} by considering only the diagonal elements of $\mathbf{Z}_{\rm{SS}}$; and (iii) $\mathbf{Z}_{D}$ is an ${{N_{{\rm{RIS}}}} \times {N_{{\rm{RIS}}}}}$ diagonal matrix that is introduced for convenience of optimization. More specifically, $\mathbf{Z}_{D}$ can be thought of as a small perturbation that is optimized, through iterative small increments, in lieu of $\mathbf{Z}_{\rm{RIS}}$. At each iteration of Algorithm \ref{alg:Nt1Nr1_coupling_overall_procedures}, the optimal $\mathbf{Z}_{D}$, i.e., $\mathbf{Z}_{D}^*$, is obtained as the solution of the optimization problem:
\begin{subequations}
\begin{align}
& \mathcal{P}_{D}^{(k)}: \; \max_{\mathbf{Z}_{D}^{(k)}} \left|Z_{\rm{RT}}-\mathbf{z}_{\rm{RS}} (\mathbf{Z}_{\rm{SS}}+\mathbf{Z}_{\rm{RIS}}^{(k)}+\mathbf{Z}_{D}^{(k)} )^{-1} \mathbf{z}_{\rm{ST}} \right| \label{eq:He2e_SISO_coupling_iterative_modi}\\
 & \mathrm{s. \, t.} \quad \; \; \mathbf{Z}_{D}^{(k)}(i,i) = \delta \exp({j\theta_i}), \quad i= 1,2,\ldots, {N_{{\rm{RIS}}}}
\end{align}
\end{subequations}
\noindent where $\delta  \ll 1$ is the absolute value of each diagonal element of $\mathbf{Z}_{D}$ and $\theta_i$ is the phase of the $i$th diagonal element.

Once $\mathbf{Z}_{D}^{(k)*}$ at the $k$th iteration is obtained by solving $\mathcal{P}_{D}^{(k)}$, the matrix of tunable impedances of the RIS at the $(k+1)$th iteration is updated as follows:
\begin{equation}
\mathbf{Z}_{\rm{RIS}}^{(k+1)}=\mathbf{Z}_{\rm{RIS}}^{(k)}+ j\textrm{Im}\left(\mathbf{Z}_{D}^{(k)*} \right) \label{eq:updation_Z_ris_k_plus_1}
\end{equation}

It is worth noting that only the imaginary part of $\mathbf{Z}_{D}^{(k)*}$ is used for updating $\mathbf{Z}_{\rm{RIS}}^{(k+1)}$, in order to ensure that the real part of each element of $\mathbf{Z}_{\rm{RIS}}$ is equal to $R_0$ (as required in $\mathcal{P}$).

In the following two subsections, we introduce an efficient solution for solving $\mathcal{P}_D^{(k)}$, provide guidelines on the choice of $\delta$, as well as prove that Algorithm \ref{alg:Nt1Nr1_coupling_overall_procedures} is convergent.

\vspace{-0.1cm}
\subsection{Solution of $\mathcal{P}_D^{(k)}$}

In order to efficiently solve $\mathcal{P}_D^{(k)}$ and to avoid the calculation of the inverse of a non-diagonal matrix, we leverage the Neuman series approximation for the inverse of matrices \cite{Neuman_series_source}. For ease of notation, we define $\mathbf{G}_{k} = \mathbf{Z}_{\rm{SS}} + \mathbf{Z}_{\rm{RIS}}^{(k)}$. Under the considered case of interest in which $\mathbf{G}_k$ is invertible, the application of the Neuman series yields the following:
\begin{align}
( \mathbf{G}_k +\mathbf{Z}_{D}^{(k)})^{-1} 
& = \left[ \mathbf{G}_k (\mathbf{I}_{{N_{{\rm{RIS}}}} \times {N_{{\rm{RIS}}}}}+{\mathbf{G}_k^{-1}}\mathbf{Z}_{D}^{(k)} )\right] ^{-1}\nonumber \\
& \hspace{-2cm} =  \left[ \mathbf{I}_{{N_{{\rm{RIS}}}} \times {N_{{\rm{RIS}}}}} - ( -{\mathbf{G}_k^{-1}} \mathbf{Z}_{D}^{(k)}  ) \right] ^{-1} {\mathbf{G}_k^{-1}} \nonumber\\
& \hspace{-2cm} = \sum_{n=0}^{+\infty} ( -{\mathbf{G}_k^{-1}} \mathbf{Z}_{D}^{(k) }  )^{n}{\mathbf{G}_k^{-1}} \mathop  \approx \limits^{\left( a \right)}  {\mathbf{G}_k^{-1}} - {\mathbf{G}_k^{-1}} \mathbf{Z}_{D}^{(k)} {\mathbf{G}_k^{-1}}
\label{eq:Neuman_series_inverse_sum_matrices_appro}
\end{align}
\noindent where (a) is obtained by retaining the first two terms ($n=0$ and $n=1$) of the Neuman series representation.

Since only two terms of the Neuman series are used in \eqref{eq:Neuman_series_inverse_sum_matrices_appro}, it is necessary to ensure that the obtained approximation is sufficiently accurate at each iteration step. To this end, an appropriate choice for $\delta$ is needed. Let $\boldsymbol \Gamma= \left[ \mathbf{I}_{{N_{{\rm{RIS}}}} \times {N_{{\rm{RIS}}}}}- (-{\mathbf{G}_k^{-1}}\mathbf{Z}_{D}^{(k)} ) \right] ^{-1}$ and $\boldsymbol{\widehat{\Gamma}}=\mathbf{I}_{{N_{{\rm{RIS}}}} \times {N_{{\rm{RIS}}}}}-{\mathbf{G}_k^{-1}}\mathbf{Z}_{D}^{(k)}$ be the exact matrix that needs to be inverted and its approximation obtained from the Neuman series. From \cite[Eq. (4.17)]{Neuman_series_source}, the norm of the error is upper bounded as follows:
\begin{equation}
    \| \boldsymbol \Gamma- \boldsymbol{\widehat{\Gamma}} \| \leq \frac{\|  ( {\mathbf{G}_k^{-1}} \mathbf{Z}_{D}^{(k)}  ) \|^{2} }{1-\| {\mathbf{G}_k^{-1}}\mathbf{Z}_{D}^{(k)} \| }
\end{equation}

Therefore, a sufficient condition for the approximation error to be small is $\| \boldsymbol \Gamma- \boldsymbol{\widehat{\Gamma}} \| \ll 1$, which yields $\| {\mathbf{G}_k^{-1}}\mathbf{Z}_{D}^{(k)} \| \ll 1$. Since $\| {\mathbf{G}_k^{-1}}\mathbf{Z}_{D}^{(k)} \| \leq \| {\mathbf{G}_k^{-1}} \| \| \mathbf{Z}_{D}^{(k)} \| \leq \| {\mathbf{G}_k^{-1}} \| \delta$, we obtain the sufficient condition $\delta \ll 1/ \| {\mathbf{G}_k^{-1}} \|$. This implies that the two-term Neuman series approximation in \eqref{eq:Neuman_series_inverse_sum_matrices_appro} is sufficiently accurate provided that the absolute value $\delta$ is sufficiently small at each iteration of Algorithm \ref{alg:Nt1Nr1_coupling_overall_procedures}.

By leveraging the Neuman series approximation in \eqref{eq:Neuman_series_inverse_sum_matrices_appro}, the optimization problem $\mathcal{P}_D^{(k)}$ can reformulated as follows:
\begin{subequations}
\begin{align}
& \mathcal{\widetilde P}_{D}^{(k)}: \; \mathop {\max }\limits_{{\bf{Z}}_D^{\left( k \right)}} \left| {{b^{\left( k \right)}} + {{\bf{p}}^{\left( k \right)}}{\bf{Z}}_D^{\left( k \right)}{{\bf{q}}^{\left( k \right)}}} \right| \label{eq:He2e_SISO_coupling_iter_final_problem}\\
 & \mathrm{s. \, t.} \quad \; \; \mathbf{Z}_{D}^{(k)}(i,i) = \delta \exp({j\theta_i}), \quad i= 1,2,\ldots, {N_{{\rm{RIS}}}}
\end{align}
\end{subequations}
\noindent where the following definitions are introduced:
\begin{align}
& {b^{\left( k \right)}} = {Z_{{\rm{RT}}}} - {{\bf{z}}_{{\rm{RS}}}}{\left( {{{\bf{Z}}_{{\rm{SS}}}} + {\bf{Z}}_{{\rm{RIS}}}^{\left( k \right)}} \right)^{ - 1}}{{\bf{z}}_{{\rm{ST}}}}\\
& {{\bf{p}}^{\left( k \right)}} = {{\bf{z}}_{{\rm{RS}}}}{\left( {{{\bf{Z}}_{{\rm{SS}}}} + {\bf{Z}}_{{\rm{RIS}}}^{\left( k \right)}} \right)^{ - 1}}\\
& {{\bf{q}}^{\left( k \right)}} = {\left( {{{\bf{Z}}_{{\rm{SS}}}} + {\bf{Z}}_{{\rm{RIS}}}^{\left( k \right)}} \right)^{ - 1}}{{\bf{z}}_{{\rm{ST}}}}
\end{align}

The following theorem yields the optimal solution of $\mathcal{\widetilde P}_{D}^{(k)}$.

\begin{theorem} \label{thm:SISO_theorem}
The global maximizer of $\mathcal{\widetilde P}_{D}^{(k)}$ is as follows (for $i= 1,2,\ldots, {N_{{\rm{RIS}}}}$):
\begin{equation}  \label{thm:OptimalZD}
    {\bf{Z}}_D^{\left( k \right)*}\left( {i,i} \right) = \delta \exp \left( {jw_{\pi}\left( {\angle {b^{\left( k \right)}} - \angle {{\bf{p}}^{\left( k \right)}}\left( i \right) - \angle {{\bf{q}}^{\left( k \right)}}\left( i \right)} \right)} \right)
\end{equation}
\end{theorem}
\begin{IEEEproof}
The proof follows by using the same steps as for the proof of Theorem \ref{thm:SISO_NC_theorem} for computing the optimal $\theta_i$.
\end{IEEEproof}

\subsection{Convergence of Algorithm \ref{alg:Nt1Nr1_coupling_overall_procedures}}

Armed with Theorem \ref{thm:SISO_theorem},  we analyze the convergence of Algorithm \ref{alg:Nt1Nr1_coupling_overall_procedures}. For ease of writing, we introduce the notation:
\begin{equation} \label{ObjectiveIteration}
{\mathcal{L}}\left( {{\bf{Z}}_{{\rm{RIS}}}^{\left( k \right)},{\bf{Z}}_D^{\left( k \right)*}} \right) = \left| {{b^{\left( k \right)}} + {{\bf{p}}^{\left( k \right)}}\left( {j{\mathop{\rm Im}\nolimits} \left( {{\bf{Z}}_D^{\left( k \right)*}} \right)} \right){{\bf{q}}^{\left( k \right)}}} \right|
\end{equation}
\noindent which is the objective function of $\mathcal{\widetilde P}_{D}^{(k)}$, at the $k$th iteration of Algorithm \ref{alg:Nt1Nr1_coupling_overall_procedures}, that is evaluated at ${\bf{Z}}_D^{\left( k \right)*}$ in \eqref{thm:OptimalZD}. In \eqref{ObjectiveIteration}, in particular, we consider only the imaginary part of ${\bf{Z}}_D^{\left( k \right)*}$, since it determines the objective function according to \eqref{eq:updation_Z_ris_k_plus_1}.

The main result about the convergence of Algorithm \ref{alg:Nt1Nr1_coupling_overall_procedures} is stated in the following proposition.

\begin{proposition} \label{prop:He2e_SISO_opt_alg_convergence}
The objective function of $\mathcal{\widetilde P}_{D}^{(k)}$ at the $k$th iteration of Algorithm \ref{alg:Nt1Nr1_coupling_overall_procedures}, i.e., ${\mathcal{L}}\left( {{\bf{Z}}_{{\rm{RIS}}}^{\left( k \right)},{\bf{Z}}_D^{\left( k \right)*}} \right)$ in \eqref{ObjectiveIteration}, is a non-decreasing function of ${{\bf{Z}}_D^{\left( k \right)*}}$ and is upper bounded. Therefore, Algorithm \ref{alg:Nt1Nr1_coupling_overall_procedures} is convergent.
\end{proposition}
\begin{IEEEproof}
To prove that ${\mathcal{L}}\left( {{\bf{Z}}_{{\rm{RIS}}}^{\left( k \right)},{\bf{Z}}_D^{\left( k \right)*}} \right)$ is non-decreasing, we need to prove ${\mathcal{L}}\left( {{\bf{Z}}_{{\rm{RIS}}}^{\left( k \right)},{\bf{Z}}_D^{\left( k \right)*}} \right) \le {\mathcal{L}}\left( {{\bf{Z}}_{{\rm{RIS}}}^{\left( {k + 1} \right)},{\bf{Z}}_D^{\left( {k + 1} \right)*}} \right)$. To this end, we have the following two results:
\begin{align}
& {\mathcal{L}}\left( {{\bf{Z}}_{{\rm{RIS}}}^{\left( {k + 1} \right)},{{\bf{0}}_{{N_{{\rm{RIS}}}} \times {N_{{\rm{RIS}}}}}}} \right) = \left| {{b^{\left( {k + 1} \right)}}} \right|\\
&\mathop  = \limits^{\left( a \right)} \left| {{Z_{{\rm{RT}}}} - {{\bf{z}}_{{\rm{RS}}}}{{\left( {{{\bf{Z}}_{{\rm{SS}}}} + {\bf{Z}}_{{\rm{RIS}}}^{\left( {k + 1} \right)}} \right)}^{ - 1}}{{\bf{z}}_{{\rm{ST}}}}} \right|\\
& \mathop  = \limits^{\left( b \right)} \left| {{Z_{{\rm{RT}}}} - {{\bf{z}}_{{\rm{RS}}}}{{\left( {{{\bf{Z}}_{{\rm{SS}}}} + {\bf{Z}}_{{\rm{RIS}}}^{\left( k \right)} + j{\mathop{\rm Im}\nolimits} \left( {{\bf{Z}}_D^{\left( k \right)*}} \right)} \right)}^{ - 1}}{{\bf{z}}_{{\rm{ST}}}}} \right|\\
& \mathop  \approx \limits^{\left( c \right)} \left| {{b^{\left( k \right)}} + {{\bf{p}}^{\left( k \right)}}\left( {j{\mathop{\rm Im}\nolimits} \left( {{\bf{Z}}_D^{\left( k \right)*}} \right)} \right){{\bf{q}}^{\left( k \right)}}} \right| \\
& = {\mathcal{L}}\left( {{\bf{Z}}_{{\rm{RIS}}}^{\left( k \right)},{\bf{Z}}_D^{\left( k \right)*}} \right)
\end{align}
\begin{align}
& {\mathcal{L}}\left( {{\bf{Z}}_{{\rm{RIS}}}^{\left( {k + 1} \right)},{\bf{Z}}_D^{\left( {k + 1} \right)*}} \right)\\
& = \left| {{b^{\left( {k + 1} \right)}} + {{\bf{p}}^{\left( {k + 1} \right)}}\left( {j{\mathop{\rm Im}\nolimits} \left( {{\bf{Z}}_D^{\left( {k + 1} \right)*}} \right)} \right){{\bf{q}}^{\left( {k + 1} \right)}}} \right|\\
& \mathop  = \limits^{\left( d \right)} \left| {{b^{\left( {k + 1} \right)}} + j\delta \sum\limits_{i = 1}^{{N_{{\rm{RIS}}}}} {a_i^{\left( {k + 1} \right)}\sin \left( {\theta _i^{\left( {k + 1} \right)}} \right)} } \right|\\
& = \left| {\left| {{b^{\left( {k + 1} \right)}}} \right| + j\delta \sum\limits_{i = 1}^{{N_{{\rm{RIS}}}}} {\left| {a_i^{\left( {k + 1} \right)}} \right|{e^{ - j\theta _i^{\left( {k + 1} \right)}}}\sin \left( {\theta _i^{\left( {k + 1} \right)}} \right)} } \right|\\
& \mathop  = \limits^{\left( e \right)} \left| {\left| {{b^{\left( {k + 1} \right)}}} \right| + {X^{\left( {k + 1} \right)}} + j{Y^{\left( {k + 1} \right)}}} \right|\\
& \mathop  \ge \limits^{\left( f \right)} \left| {{b^{\left( {k + 1} \right)}}} \right| = {\mathcal{L}}\left( {{\bf{Z}}_{{\rm{RIS}}}^{\left( {k + 1} \right)},{{\bf{0}}_{{N_{{\rm{RIS}}}} \times {N_{{\rm{RIS}}}}}}} \right)
\end{align}
\noindent where (a) follows by definition of $b^{(k)}$; (b) follows from \eqref{eq:updation_Z_ris_k_plus_1}; (c) follows by applying the Neuman series approximation; (d) follows by using the notation $a_i^{\left( {k + 1} \right)} = {{\bf{p}}^{\left( {k + 1} \right)}}\left( i \right){{\bf{q}}^{\left( {k + 1} \right)}}\left( i \right)$ and $\theta _i^{\left( {k + 1} \right)} = w_{\pi} \left( \angle {b^{\left( {k + 1} \right)}} - \angle {{\bf{p}}^{\left( {k + 1} \right)}}\left( i \right) - \angle {{\bf{q}}^{\left( {k + 1} \right)}}\left( i \right)\right)$; (e) follows by introducing the notation ${X^{\left( {k + 1} \right)}} = \delta \sum\nolimits_{i = 1}^{{N_{{\rm{RIS}}}}} {\left| {a_i^{\left( {k + 1} \right)}} \right|{{\sin }^2}\left( {\theta _i^{\left( {k + 1} \right)}} \right)}$ and ${Y^{\left( {k + 1} \right)}} = \delta \sum\nolimits_{i = 1}^{{N_{{\rm{RIS}}}}} {\left| {a_i^{\left( {k + 1} \right)}} \right|\sin \left( {\theta _i^{\left( {k + 1} \right)}} \right)\cos \left( {\theta _i^{\left( {k + 1} \right)}} \right)}$; and (f) follows because ${X^{\left( {k + 1} \right)}} \ge 0$ based on its definition.

Therefore, we eventually obtain ${\mathcal{L}}\left( {{\bf{Z}}_{{\rm{RIS}}}^{\left( {k + 1} \right)},{\bf{Z}}_D^{\left( {k + 1} \right)*}} \right) \ge {\mathcal{L}}\left( {{\bf{Z}}_{{\rm{RIS}}}^{\left( {k + 1} \right)},{{\bf{0}}_{{N_{{\rm{RIS}}}} \times {N_{{\rm{RIS}}}}}}} \right) \approx {\mathcal{L}}\left( {{\bf{Z}}_{{\rm{RIS}}}^{\left( k \right)},{\bf{Z}}_D^{\left( k \right)*}} \right)$, which proves that the objective function is non-decreasing.

To prove that ${\mathcal{L}}\left( {{\bf{Z}}_{{\rm{RIS}}}^{\left( k \right)},{\bf{Z}}_D^{\left( k \right)*}} \right)$ is upper bounded, we can first apply the triangle inequality to the objective function in \eqref{eq:He2e_SISO_coupling_iterative_modi} and then consider that the matrix  $\mathbf{G}_{k} = \mathbf{Z}_{\rm{SS}} + \mathbf{Z}_{\rm{RIS}}^{(k)}$ is invertible in the case study of interest. This implies that each element of the inverse of $\mathbf{G}_{k}$ can be upper bounded by the largest of its elements. This concludes the proof.
\end{IEEEproof}
\begin{figure}[!t]
\centering
\includegraphics[width=0.80\columnwidth]{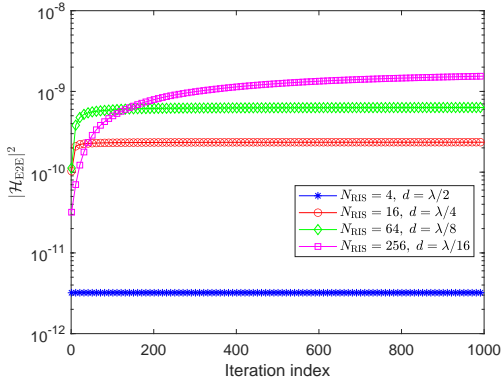}
\vspace{-0.10cm} \caption{Convergence of Algorithm \ref{alg:Nt1Nr1_coupling_overall_procedures}.}
\label{Fig_1} \vspace{-0.5cm}
\end{figure}
\begin{figure}[!t]
\centering
\includegraphics[width=0.80\columnwidth]{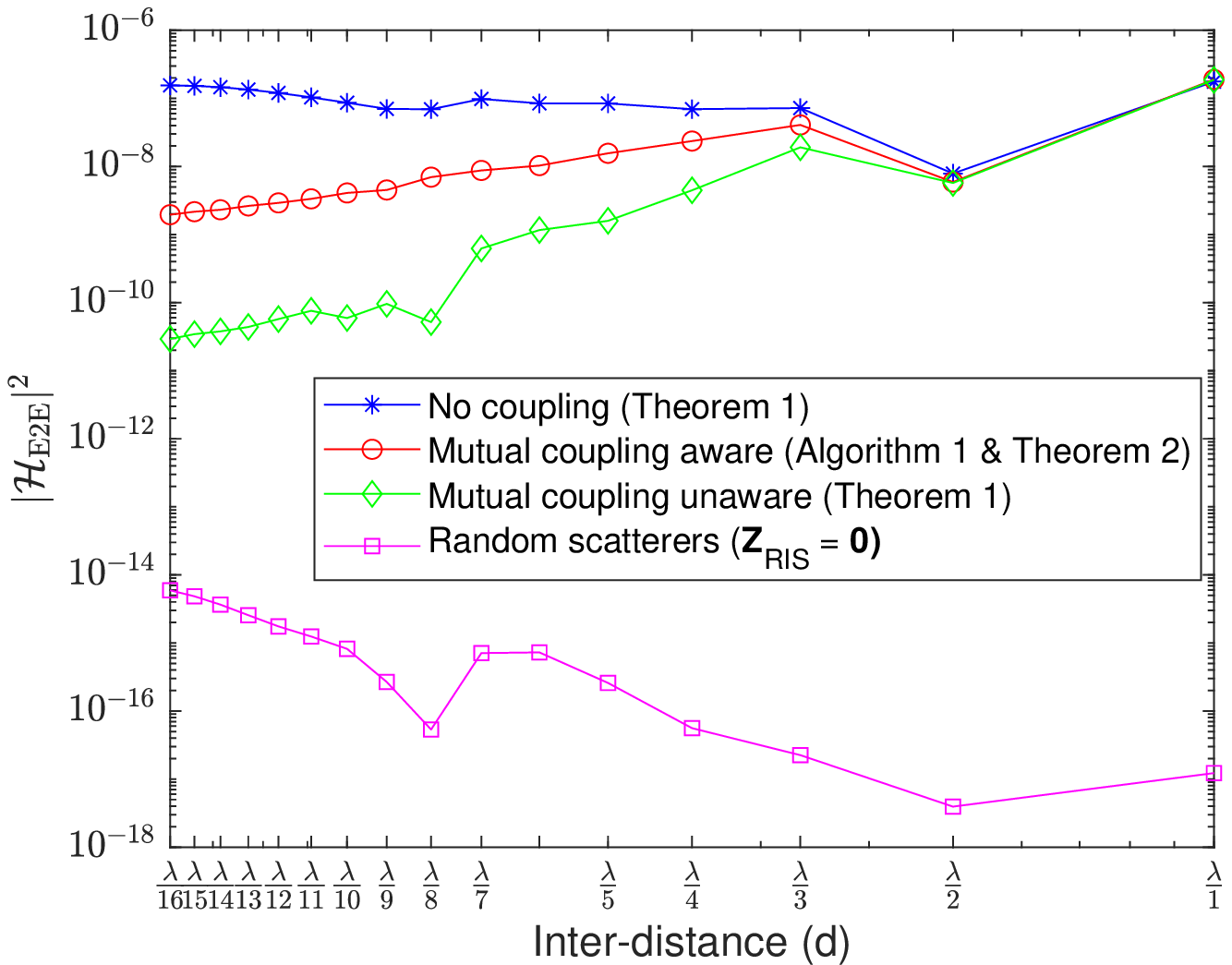}
\vspace{-0.10cm} \caption{Impact of mutual coupling (${N_{{\rm{RIS}}}}$ is kept fixed).}
\label{Fig_2} \vspace{-0.5cm}
\end{figure}
\begin{figure}[!t]
\centering
\includegraphics[width=0.80\columnwidth]{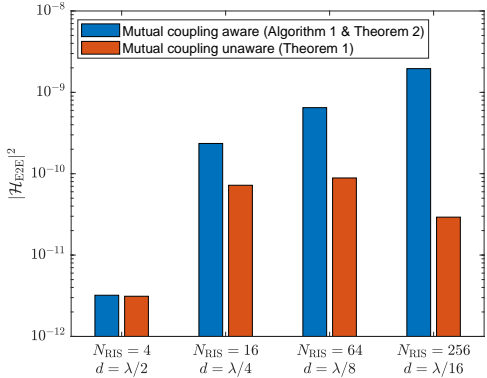}
\vspace{-0.10cm} \caption{Impact of mutual coupling (the RIS size is kept fixed).}
\label{Fig_3} \vspace{-0.5cm}
\end{figure}
\section{Numerical Results} \label{Numerical_Results} 
We illustrate some numerical results in order to verify the effectiveness of the proposed optimization algorithms and to analyze the impact of mutual coupling. The considered setup is the same as in \cite{Gabriele}: the carrier frequency is $f= 28$ GHz; the locations of transmitter and receiver are ${{\bf{r}}_{t}} = (5, -5, 3)$ and ${{\bf{r}}_{r}} = (5, 5, 1)$; and the RIS is centered at $(0,0,0)$ with $M=\sqrt{ N_{\rm{RIS}} }$. The transmit and receive antennas, and the passive scatterers of the RIS are identical thin wires with radius $a=\lambda/500$ and length $l=\lambda/32$, where $\lambda$ is the wavelength, and ${R_0} = 0.2$ Ohm. We assume that the direct link is negligible due to the presence of obstacles between the transmitter and receiver. The self and mutual impedances (${{\boldsymbol{{\rm Z}}}_{{\rm{SS}}}}$) are obtained from the analytical framework in \cite[Lemma 2]{Gabriele}.

In Fig. \ref{Fig_1}, we verify the convergence of Algorithm \ref{alg:Nt1Nr1_coupling_overall_procedures} as a function of the number of scattering elements of the RIS ($N_{\rm{RIS}}$) and their inter-distance ($d$). We observe that the statement in Proposition \ref{prop:He2e_SISO_opt_alg_convergence} is confirmed. The convergence rate of Algorithm \ref{alg:Nt1Nr1_coupling_overall_procedures} depends, in general, on both $N_{\rm{RIS}}$ and $d$.

In Fig. \ref{Fig_2}, we analyze the impact of mutual coupling on the end-to-end SNR and the effectiveness of mutual coupling aware optimization. The figure is obtained by keeping fixed $N_{\rm{RIS}}$, while varying the inter-distance $d$ between adjacent scattering elements. The case study corresponding to ``no coupling'' is referred to a setup in which only the diagonal elements of ${{\boldsymbol{{\rm Z}}}_{{\rm{SS}}}}$ are retained (ideal case study). The case study corresponding to ``mutual coupling unaware'' is referred to a setup in which the off-diagonal elements of ${{\boldsymbol{{\rm Z}}}_{{\rm{SS}}}}$ are non-zero, but they are not taken into account for system optimization (mismatched optimization). We observe that the proposed mutual coupling aware design and algorithm can significantly enhance the intensity of the end-to-end SNR.

In Fig. \ref{Fig_3}, we analyze the impact of mutual coupling by assuming that the size of the RIS is kept fixed (i.e., ${N_{{\rm{RIS}}}}{d^2}$ is constant), while $N_{\rm{RIS}}$ and $d$ are varied accordingly. This case study corresponds to the optimization of one super-cell of size $\lambda \times \lambda$ (see \cite[Fig. 4]{MDR_JSAC}). We evince that ignoring the mutual coupling among the scattering elements of the RIS results in a performance degradation, as compared to the proposed mutual coupling aware optimization (i.e., Algorithm \ref{alg:Nt1Nr1_coupling_overall_procedures}).

\section{Conclusion} \label{Conclusion} 
By leveraging a recently proposed communication model for RIS-assisted communications, we have introduced an optimization algorithm for maximizing the end-to-end SNR as a function of the tunable impedances of the RIS, which explicitly takes into account the mutual coupling among closely spaced scattering elements. The numerical results have unveiled that the end-to-end SNR can be enhanced by explicitly taking into account the mutual coupling at the design phase. The proposed approach is applicable to single-antenna transmitters and receivers that operate in the far-field of the RIS. Possible generalization of the proposed approach includes the optimization of RIS-assisted systems with multi-antenna transmitters and receivers in both the far-field and near-field regimes.

\bibliographystyle{IEEEtran}

\end{document}